\documentclass[apl,twocolumn,preprintnumbers,amsmath,amssymb,floatfix]{revtex4-2}
\usepackage{graphicx}% Include figure files
\usepackage{lmodern}
\usepackage{tabularx}
\usepackage{siunitx}
\usepackage{booktabs}
\usepackage{etoolbox}
\usepackage{epstopdf}
\usepackage{float}
\usepackage{placeins}
\usepackage{hyperref,graphicx}
\usepackage{dcolumn}% Align table columns on decimal point
\usepackage{bm}% bold math
\usepackage{color}
\usepackage{natbib}
\usepackage{amsmath}
\usepackage{amssymb}
\usepackage{multirow} % for row & col SPanning in tables
\usepackage{hyperref}
\usepackage{stackengine}
\usepackage{xcolor}
\DeclareUnicodeCharacter{0308}{\"{}}
\hypersetup{colorlinks=true, linkcolor=blue, citecolor=red, urlcolor=magenta, pdftitle={DDPOs}, pdfauthor={Saurabh Ghosh}}

\usepackage[normalem]{ulem}

\begin{document}
%%%%%%%%%%%%%%%%%%%%%%%%%%%%%%%%
%\title{First-principles study of ferro/ferrimagnetic insulating double-double perovskite oxides} 
\title{Ferroelectric metal-organic frameworks as wide band gap materials} %for Spintronic and Optical Device Applications}
%\title{Unveiling Insulating Ferro and Ferrimagnetism in Double-Double Perovskite Oxides}
%\title{First principles investigation on electronic, magnetic and optical properties of double-double perovskite oxides} 
\author{Monirul Shaikh$^{\dagger, \ddagger}$}
\email{msk.phe@gmail.com}
\author{Sathiyamoorthy Buvaneswaran$^{\dagger}$}
\author{Asif Latief Bhat$^{\dagger}$}
\author{Trilochan Sahoo$^{\dagger}$}
\author{Saurabh Ghosh$^{\dagger, \ddagger}$}
\email{saurabhghosh2802@gmail.com}
\affiliation{$^{\dagger}$ Department of Physics and Nanotechnology, SRM Institute of Science and Technology, Kattankulathur - 603 203, Tamil Nadu, India}
\affiliation{$^{\ddagger}$ Center for Advanced Computational and Theoretical Sciences, SRM Institute of Science and Technology, Kattankulathur - 603 203, Tamil Nadu, India}
%%%%%%%%%%%%%%%%%%%%%%%%%%%%%
\begin{abstract}
Wide band gap materials are particularly relevant at high temperatures. 
The band gap shrinkage at higher temperatures prevents device applications with narrow band gap semiconductors. 
Considering $\alpha$-phase strontium cyanurate as a prototype structure, we identify a group of metal-organic frameworks (MOFs) that exhibit ultra-wide band gaps ranging from 5.5 to 5.7 eV.
Recently, a strontium cyanurate compound was found to undergo a phase transition from a high-symmetry $\beta$-phase to a low-symmetry ferroelectric $\alpha$-phase when the temperature was reduced.
In the present study, utilizing group theory techniques, we unravel that a zone-center $\Gamma_2^-$ phonon mode modifies our structures from high-symmetry $\beta$-phase to a low-symmetry $\alpha$-phase for A$_3$(O$_3$C$_3$N$_3$)$_2$ MOFs with A = Mg, Ca, Sr, and Ba.
We implement first-principles calculations to investigate structural, ferroelectric, and optical properties of these compounds in $\alpha$-phase.
\textcolor{black}{The switching barriers between bistable polar states are also estimated.} 
\textcolor{black}{Further, to realize their feasibility, we examine the dynamical and \textcolor{black}{thermal} stabilities for all of these MOFs.}
\end{abstract}
\maketitle
%%%%%%%%%%%%%%%%%%%%%%%%%%%%%
\section{Introduction}
Wide band gap (WBG) materials are crucial in high temperature, deep-ultraviolet detectors (DUV), and in power electronic devices \cite{WBG_Ga2O3, WBG_GaN}. However, wide or ultra-wide semiconductors are rare \cite{WBG_GaN, WBG_AlGaN}. Up to now, WBG semiconductors have been fundamentally observed in toxic lead-based inorganic, organic-inorganic type perovskites \cite{liao2015lead, li2022cspbcl3, li2024generic, grandhi2024wide} and in gallium-based oxides and nitrides \cite{WBG_GaN, WBG_AlGaN}. WBG materials or WBG semiconductors are often called materials of the future due to their scarcity and expensive growth techniques \cite{WBG_CuAlO2, woods2020wide}. The WBG materials become superior when they exhibit ferroelectricity \cite{WBG_perovskite} due to the absence of free charge carriers. Ferroelectricity possesses a spontaneous polarization at the Curie temperature (T$_c$), and the polarization can be reversed by an applied electric field. Therefore, in the crystalline unit cell, ferroelectrics possess a permanent electric dipole and also ferroelectrics show hysteresis \cite{asadi2016ferroelectricity}. Due to their electric dipoles and hence spontaneous polarization, the ferroelectric semiconductors are crucial for non-volatile memory and energy harvesting applications \cite{yu2022hardware, colombo2024future, wang2024perspectives}. A wide band gap together with ferroelectricity is of particular interest to prevent electron transport from valence band maxima to conduction band minima to efficiently utilize them for high temperature and power electronics. Keeping the sustainable environment in mind, we choose to work with wide band gap materials within the metal-organic frameworks (MOFs) family that exhibit ferroelectricity. 
\par
\textcolor{black}{Ferroelectric-based MOFs have garnered significant attention for their multifunctional \cite{xian20252d, wang2025enhancing, yang2025large, kitou2025piezoelectric, guerin2025expanding, song2025ferroelectric, chen2025colossal} and multiferroic \cite{zhou2025pressure, cheng2025multiferroicity} applications. Extensive experimental work has demonstrated the rich chemistry and functionality of MOFs \cite{zhang2012ferroelectric, pan2014resistance, li2018ba}. 
MOFs are hybrid crystalline compounds comprised of an extended ordered network of organic linkers and metal cations, giving rise to highly designable frameworks with ultra-low densities, good thermal stability, and discrete, well-defined coordination environments. 
These characteristics have enabled a broad range of applications, including gas storage, separation, catalysis, sensing, and drug delivery and catalysis, among other applications \cite{gangu2016review, schoedel2016structures, zhou2014experimental, asadi2016ferroelectricity, reinsch2013structures}. In addition, ferroelectric MOFs have been explored in metal–ferroelectric–semiconductor (MFS) architectures as promising candidates for nonvolatile memory devices, further motivating efforts to design and understand ferroelectric MOFs. In parallel, the realization of switchable polarization in MOFs has opened opportunities for integrating them into ferroelectric and multiferroic devices, such as metal–ferroelectric semiconductor (MFS) structures for nonvolatile memory and logic. 
The key challenge is to identify frameworks that combine robust ferroelectric order with sufficient chemical and thermal stability and, ideally, wide band gaps to suppress leakage at elevated temperatures. 
Against this backdrop, exploiting ferroelectric MOFs with controllable structure–property relationships and wide band gaps is the central focus of the present work.}

\textcolor{black}{On the theoretical side, there have been substantial advances in understanding structure, dynamics, and ferroic order in MOFs \cite{kolesnikov2019metal, schmid2017electric, ghoufi2017electrically, dürholt2019tuning, ghosh2015strain, shaikh2022defect, gu2025ferroelectric}. 
Atomistic simulations and symmetry-based analyses have clarified how framework flexibility, including linker rotations, metal–ligand coordination changes, and breathing distortions, can couple to polarization degrees of freedom and drive structural phase transitions. 
In particular, the coupling between breathing distortion and ferroelectricity, and the associated bistable energy landscape, has been extensively investigated \cite{kolesnikov2019metal, schmid2017electric, ghoufi2017electrically}. 
These studies have established that relatively small structural rearrangements can lead to large changes in dipole alignment, thereby enabling switchable polarization in otherwise non-traditional ferroelectrics. 
Furthermore, Rochus Schmid and co-workers have demonstrated that external electric fields can act as an efficient stimulus to trigger breathing and other structural changes in MOFs \cite{dürholt2019tuning}, providing a microscopic basis for electric-field control of framework topology. 
Most recently, theoretical work has revealed ferroelectrically switchable altermagnetism in MOFs \cite{gu2025ferroelectric}, underscoring the rich interplay between spin, charge, and lattice degrees of freedom. 
Taken together, these developments provide a robust theoretical framework for describing structural transitions in MOFs and their coupling to ferroelectricity, and they motivate a systematic exploration of new MOF families where polarization, switching barriers, and band gaps can be co-optimized.}

\textcolor{black}{Interestingly, a subset of ferroelectrics exhibits wide band gaps and, consequently, attractive optical properties \cite{li2018ba}. 
In several of these systems, the presence of a planar (C$_3$N$_3$O$_3$)$^{3-}$ group has been implicated as a key structural motif governing both the electronic structure and the dielectric response. 
Kalmutzki \textit{et al.} reported that SrCNO, which contains such a planar cyanurate unit, undergoes a temperature-induced phase transition from a high-symmetry (\textit{R3c}) $\beta$-phase to a low-symmetry (\textit{Cc}) $\alpha$-phase upon cooling \cite{kalmutzki2017formation}, thereby establishing a direct link between framework distortions and ferroelectric behavior. 
Motivated by these observations, and by the need for wide-band-gap ferroelectric materials capable of operating at high temperatures, we systematically investigate cyanurate-based MOFs of the form A$_3$(O$_3$C$_3$N$_3$)$_2$ (A = Mg, Ca, Sr, Ba) in their $\alpha$-phase. 
By ``clubbing’’ this family with the cyanurates containing (C$_3$N$_3$O$_3$)$^{3-}$ group, we are able to elucidate how the A-site cation size and chemistry tune the structural distortions, spontaneous polarization, switching barriers, and optical band gaps within a single chemically coherent series. 
Our investigation, therefore, not only extends the current state of the art on ferroelectric MOFs, as highlighted by the recent theoretical and experimental works cited above, but also provides design principles for engineering wide-band-gap ferroelectric MOFs optimized for high-temperature and optoelectronic applications.}

Considering a recently synthesized strontium cyanurate structure \cite{kalmutzki2017formation}, Sr$_3$(O$_3$C$_3$N$_3$)$_2$ (SrCNO), we identify a set of wide-band-gap ferroelectric MOFs. 
In our pursuit, we utilize modern group theory techniques to investigate the underlying principle that dictates phase transitions.
We implement first-principles density functional theory (DFT) \cite{kohn1996density} calculations to study structural, ferroelectric, and optical properties of these compounds in $\alpha$-phase.
\textcolor{black}{We also explore the switching barriers between bistable polar states. Further, we perform \textit{ab initio} molecular dynamics (AIMD) simulations and phonon calculations to comment on their thermal and dynamical stabilities, respectively.} 
%%%%%%%%%%%%%%%%%%%%%%%%%%%%%%%%%%%%%%%%%%%%%%%%%%%%%%%%%%%%%%%%%%%%%%%%%%%%%
\section{Methodology}
The first-principles density functional theory (DFT) computations are performed in steps. First, we perform geometry relaxations for structural and ferroelectric information of these compounds by a plane wave pseudo-potential code \cite{kresse1996efficient}. In this step, we carried out the calculations with the projector augmented wave (PAW) method \cite{kohn1965self} and GGA exchange-correlation functional. The convergence in total energy and Hellman-Feynman forces is set to 1 $\mu eV$ and 0.01 $eV$/$\text{Å}$ for individual atoms, respectively. The relaxation calculations are performed with $\Gamma$-centered 4$\times$4$\times$4  K-mesh with a 600.0 eV energy cut-off. The convergences are carefully tested with a higher energy cut-off and finer K-mesh density. The ferroelectric polarization is calculated using the Berry phase method \cite{king1994first} as implemented in VASP. Our study utilizes modern group theory techniques through the ISODISTORT code \cite{campbell2006isodisplace}. To analyze the geometry of our compounds, we employ the visualization for electronic and structural analysis (VESTA) software \cite{momma2011vesta}. 

Next, we perform phonon computation using the phonopy code and finite difference method to examine dynamical stability and element-resolved phonon density of states as implemented in VASP \cite{finite1, phonopy-phono3py-JPCM}.
%%%%%%%%%%%%%%%%%%%%%%%%%%%%%%%%
\begin{figure}
\centering
%\subfloat
{\includegraphics[width=\linewidth]{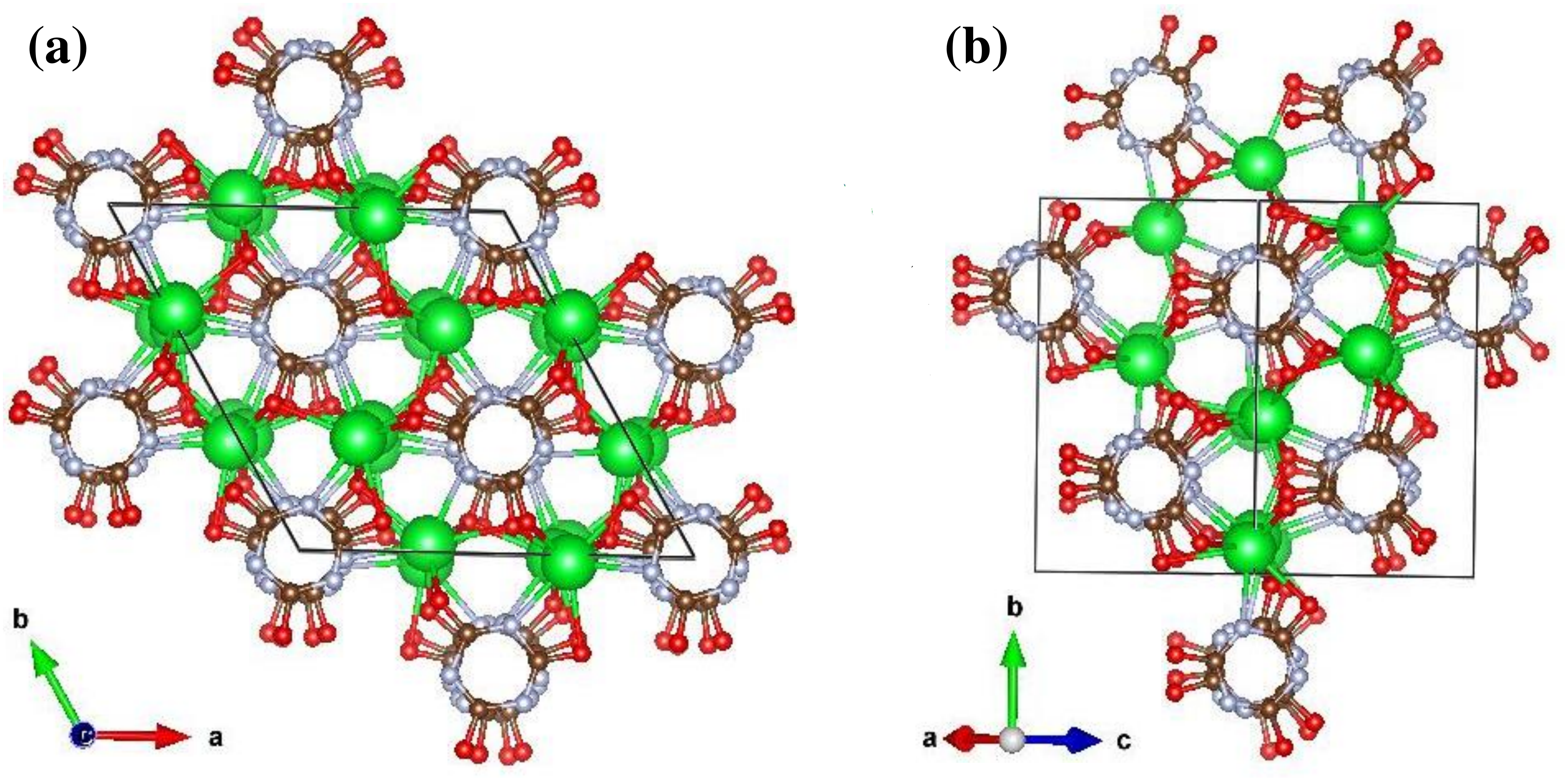}\vspace{-0pt}
} 
{\caption {\textbf{Crystal structures of the parent SrCNO compound}. (a) High symmetry trigonal $\beta$-phase (\textit{R3c}) and (b) low symmetry monoclinic $\alpha$-phase (\textit{Cc}) of Sr$_3$(O$_3$C$_3$N$_3$)$_2$ MOF. Green, brown, red, and silver-colored spheres represent
 strontium, carbon, oxygen, and nitrogen atoms, respectively.}} 
\label{Figure1}
\end{figure}
%%%%%%%%%%%%%%%%%%%%%%%%%%%%%%
\begin{table*}
\begin{center}
\setlength{\tabcolsep}{19pt}
\caption{\label{tab:1} Crystal structure information of $\alpha$-phase ACNOs}

\begin{tabular}{c|cccc|ccc}
\hline
\hline
        & \multicolumn{3}{c}{Lattice parameters}  & Cell & \multicolumn{3}{c}{Axial angles} \\
   Systems &  $a$ (\(\text{\AA}\)) & $b$ (\(\text{\AA}\)) & $c$(\(\text{\AA}\)) & Volume (\(\text{\AA}^3\)) & $\alpha$ ($^0$) & $\beta$ ($^0$) & $\gamma$ ($^0$) \\
\hline      
MgCNO & 10.63  & 10.82 & 7.73 & 888.63 & 90.0 & 91.2 & 90.0\\
CaCNO & 10.97  & 11.83 & 8.31 & 1071.99 & 90.0 & 96.5 & 90.0\\
SrCNO & 10.97  & 11.83 & 8.31 & 1071.89 & 90.0 & 96.5 & 90.0\\
BaCNO & 11.22  & 12.04 & 8.64 & 1161.67 & 90.0 & 96.0 & 90.0\\
\hline
\hline
\end{tabular}
\end{center}
\end{table*}
%%%%%%%%%%%%%%%%%%%%%%%%
\textcolor{black}{Then, the bistable polar states, as well as the switching path between them in these ferroelectric compounds, are estimated using the state-of-the-art evolution of total energy as a function of structural distortions from the non-polar structure, $C2/c$, to the polar ($Cc$) structures for all these compounds.}

\textcolor{black}{To comment on the thermal stability of these compounds, we implement \textit{ab initio} molecular dynamics (AIMD) simulations.  
The Brillouin zone is sampled at a $\Gamma$-centered K-point mesh with a 600.0 eV energy cut-off for all AIMD simulations.
The canonical NVT ensemble with the Nose-Hoover thermostat approach is utilized to keep the temperature constant during AIMD simulations\cite{nose2002molecular}.
We perform a series of AIMD simulations for temperatures of 300K, 500K, and 700K for all these systems.
All systems are allowed to relax for 10 ps with a time step of 1 fs.
}
 
Lastly, we study electronic structure and optical characteristics using a full-potential linear augmented plane-wave code \cite{schwarz2002electronic}. We utilize the modified Becke-Johnson (MBJ) potential calculation to determine the accurate band gap. The electronic and optical properties are calculated using a $\Gamma$-centered 12$\times$12$\times$12 K-mesh within the GGA+MBJ function.
%%%%%%%%%%%%%%%%
\section{Results and discussion}
%%%%%%%%%%%%%%%%
\subsection{Structural and ferroelectric properties}
%%%%%%%%%%%%%%%%%%%%%%%%%
\textbf{$\beta$-phase Sr$_3$(O$_3$C$_3$N$_3$)$_2$} : The crystal structures of parent Sr$_3$(O$_3$C$_3$N$_3$)$_2$ (SrCNO) MOF for both $\beta$-phase and $\alpha$-phase are shown in Figure 1. $\beta$-phase SrCNO crystalizes in a trigonal noncentrosymmetric space group $R3c$. The relaxed lattice parameters were found to be $a$ = 11.86 \text{\AA}, and $c$ = 12.69 \text{\AA}. $\beta$-phase SrCNO comprised of one type of strontium ion (Sr$^{2+}$) with six oxido, ($\bar{d}_{Sr1-O}$ = 2.65 \text{\AA}) and two nitrido ($\bar{d}_{Sr1-N}$ = 2.60 \text{\AA}) neighbors leading to a prism like coordination. While cyanurate ion (O$_3$C$_3$N$_3$)$^{3-}$ contains two types of carbon, nitrogen, and oxygen atoms, respectively. In the crystal structure, the cyanurate ion shows a slightly offset coplanar arrangement that leads to stacked rings in columns, as shown in Figure 1a. The $\beta$-phase is used only as a reference structure to discuss $\alpha$-phases, and hereafter we will only focus on $\alpha$-phases.

%%%%%%%%%%%%%%%%%%%%%%%%%%%%%%%%%%%%%%%%%%%%%%%%%%%%%%%%%%%%%%%%%%%%%%%%%%%%%%
\begin{table*}
\begin{center}
\setlength{\tabcolsep}{15pt}
\caption{\label{tab:2} Coordination and neighbors analysis of A-cations in $\alpha$-phase ACNOs}
\begin{tabular}{c|cccccc}
\hline
\hline
  Systems  & $\bar{d}_{A1-N}$ (\(\text{\AA}\)) & $\bar{d}_{A1-O}$ (\(\text{\AA}\)) & $\bar{d}_{A2-N}$ (\(\text{\AA}\)) & $\bar{d}_{A2-O}$ (\(\text{\AA}\)) & $\bar{d}_{A3-N}$ (\(\text{\AA}\)) & $\bar{d}_{A3-O}$ (\(\text{\AA}\))\\
    \hline  
MgCNO & 2.19 & 1.73 & 2.12 & 2.00 & 2.23 & 2.11 \\ 
CaCNO & 2.57 & 2.53 & 2.70 & 2.49 & 2.63 & 2.51 \\
SrCNO & 2.62 & 2.66 & 2.64 & 2.71 & 2.65 & 2.68 \\
BaCNO & 2.76 & 2.77 & 2.77 & 2.89 & 2.79 & 2.81 \\
\hline
\hline
\end{tabular}
\end{center}
\end{table*}
%%%%%%%%%%%%%%%%%%%%%%%%
\textbf{$\alpha$-phase Sr$_3$(O$_3$C$_3$N$_3$)$_2$}$ : \alpha$-phase SrCNO crystalizes in a monoclinic noncentrosymmetric space group $Cc$. The relaxed crystal structure information is shown in TABLE \ref{tab:1}. Contrary to the $\beta$-phase, $\alpha$-phase SrCNO contains three types of strontium ions (Sr$^{2+}$). Sr1$^{2+}$ shows five oxido, and two nitrido, while Sr2$^{2+}$ shows four oxido and two nitrido, and Sr3$^{2+}$ displays six oxido, and two nitrido neighbors. The average distances between strontium ions, and nitrogen $\&$ oxygen atoms are shown in Table \ref{tab:2}. The cyanurate ion (O$_3$C$_3$N$_3$)$^{3-}$ contains six types of carbon, nitrogen, and oxygen atoms, respectively. However, in the crystal structure, the cyanurate ion exhibits slightly offset coplanar arrangement analogous to $\beta$-phase as shown in Figure 1b. 

%%%%%%%%%%%%%%%%%%%%%%%%%%%%%%
\textbf{$\alpha$-phase Mg$_3$(O$_3$C$_3$N$_3$)$_2$}$ : \alpha$-phase MgCNO also crystalizes in a monoclinic noncentrosymmetric space group $Cc$ similar to SrCNO. The relaxed lattice parameters are significantly shorter as compared to SrCNO due to smaller magnesium ions as shown in TABLE \ref{tab:1}. MgCNO contains three types of magnesium ions (Mg$^{2+}$). Mg1$^{2+}$ shows four oxido, and two nitrido while Mg2$^{2+}$ shows two oxido and two nitrido and Mg3$^{2+}$ exhibits four oxido, and one nitrido neighbors.  The cyanurate ion (O$_3$C$_3$N$_3$)$^{3-}$ shows six types of carbon, nitrogen, and oxygen atoms respectively similar to SrCNO. However, in the crystal structure, the cyanurate ion exhibits more offset coplanar arrangement as compared to SrCNO.

\textbf{$\alpha$-phase Ca$_3$(O$_3$C$_3$N$_3$)$_2$}$ : \alpha$-phase CaCNO also crystalizes in a monoclinic noncentrosymmetric space group $Cc$ similar to SrCNO. The relaxed lattice parameters too are more or less similar to SrCNO as shown in TABLE \ref{tab:1}. CaCNO also shows three types of calcium ions (Ca$^{2+}$). Ca1$^{2+}$ shows three oxido, and two nitrido while Ca2$^{2+}$ and Ca3$^{2+}$ both show four oxido and two nitrido. The cyanurate ion (O$_3$C$_3$N$_3$)$^{3-}$ shows six types of carbon, nitrogen, and oxygen atoms respectively with similar arrangement to SrCNO.

\textbf{$\alpha$-phase Ba$_3$(O$_3$C$_3$N$_3$)$_2$}$ : \alpha$-phase BaCNO too crystalizes in a monoclinic noncentrosymmetric space group $Cc$. The relaxed lattice parameters are significantly larger as compared to SrCNO due to larger barium ions as shown in TABLE \ref{tab:1}. BaCNO also shows three types of barium ions (Ba$^{2+}$). Ba1$^{2+}$ shows five oxido, and three nitrido while Ba2$^{2+}$ and Ba3$^{2+}$ both show six oxido and two nitrido. The cyanurate ion (O$_3$C$_3$N$_3$)$^{3-}$ shows six types of carbon, nitrogen, and oxygen atoms respectively with similar arrangement to SrCNO. The average distances between A$^{2+}$, and nitrogen $\&$ oxygen atoms are shown in Table \ref{tab:2}.
%%%%%%%%%%%%%%%%%%%%%%
\begin{table}[b]
\begin{center}
\caption{\label{tab:3} \textcolor{black}{\textbf{Ferroelectric properties.} Polarization switching barriers and the values of ferroelectric polarization, \textbf{P}.}}

\begin{tabular}{c|cc}
\hline
\hline
Systems & Energy differences & \textbf{P}($\mu C/cm^2$) \\
        & $\Delta E = E_{Cc} - E_{C2/c}$ (meV/f.u.) & \\
\hline      
MgCNO & -152.1 & 1.15\\
CaCNO & -138.6 & 2.20\\
SrCNO & -69.9 & 2.39\\
BaCNO & -81.5 & 3.13\\
\hline
\hline
\end{tabular}
\end{center}
\end{table}
%%%%%%%%%%%%%%%%%%%%%%%%

\textcolor{black}{All $\alpha$-ACNO MOFs crystallize in the polar $Cc$ space group (No. 9 in International Tables for Crystallography). Since the ferroelectricity mainly originates from the A-site cations, the calculated spontaneous polarization increases with increasing ionic radius of A (see Table \ref{tab:3}). The energy barrier associated with polarization reversal is a key quantity for assessing ferroelectric switching. Direct application of the nudged elastic band (NEB) method to these MOF structures is, however, computationally demanding. Therefore, following established approaches in the MOF literature \cite{stroppa2013hybrid, jain2016switchable, li2020two}, we construct a centrosymmetric non-polar $C2/c$ structure of SrCNO and use it as the reference phase. The inset of Figure S1 in the Supplementary Materials \cite{WB-FE-MOF} shows this non-polar $C2/c$ structure (No. 15), which possesses a center of symmetry. We then identify the structural distortion that connects this non-polar phase to the polar $Cc$ phase.}
%%%%%%%%%%%%%%%%%%%%%%%%%%%%%%%%%%%%%%%%%%%%%%%%%%%%%%%%%%%%%b%%%%%%%%%%%%%%%%%

\textcolor{black}{The energy profile between the bistable polar states of SrCNO is evaluated as a prototype system by monitoring the total energy as a function of the amplitude of the symmetry-adapted polar distortion that continuously transforms the non-polar $C2/c$ structure into the polar $Cc$ structure, as shown in Figure S1 in the Supplementary Materials \cite{WB-FE-MOF}. This state-of-the-art approach is widely used to estimate switching barriers in MOFs \cite{stroppa2013hybrid, jain2016switchable, li2020two}.
The resulting double-well energy landscape between the two symmetry-related polar states confirms the ferroelectric nature of these MOFs. Table \ref{tab:3} reports the calculated energy differences between the non-polar $C2/c$ phase and the polar $Cc$ phase for all $\alpha$-ACNO MOFs, demonstrating that the ferroelectric phase is energetically favored over the centrosymmetric structure. Our results further show that the switching barriers for these compounds are lower than those reported for other ferroelectric MOFs in the literature \cite{stroppa2013hybrid, jain2016switchable, li2020two}, indicating that the polarizations should indeed be switchable.}

 %%%%%%%%%%%%%%%%%%%%%%
 \begin{figure*}
\centering
%\subfloat
{\includegraphics[width=0.99\linewidth]{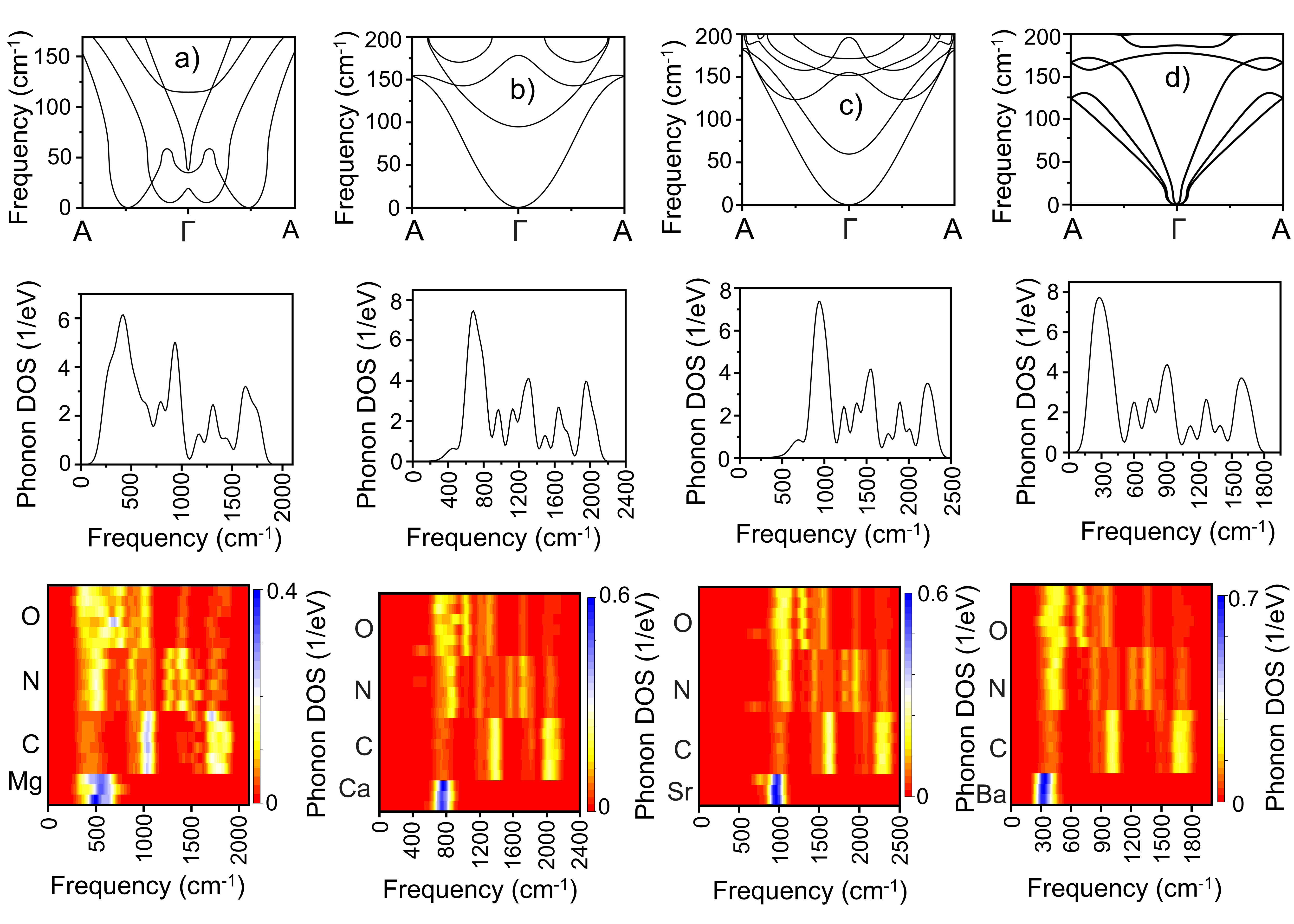}\vspace{-0pt}
} 
{\caption {\textbf{Phonon band structures and density of states}. Phonon dispersion relations for a) MgCNO, b) CaCNO, c) SrCNO, and d) BaCNO are shown in the top panel. Total density of states and element-resolved density of states for them are shown in the middle and bottom panels, respectively.} 
\label{Figure2}}
\end{figure*}
%%%%%%%%%%%%%%%%%%%%%%%%%
\subsection{Dynamical and thermal stabilities of $\alpha$-ACNOs}
%%%%%%%%%%%%%%%%%%%%%%%%%%%%%%%%%%%%%%%%%%%%%%%%%%%%%%%%%%%%%%%%%%%%%%%%%%%%%%
To identify a modification in the ferroelectricity between two phases, we decompose the high symmetry $\beta$-phase to the low symmetry $\alpha$-phase SrCNO utilizing ISODISTORT software \cite{campbell2006isodisplace}. In the process, we observe that a soft phonon $\Gamma_2^-$ mode modifies the structure to the $\alpha$-phase SrCNO. This occurs mainly due to the displacement of Sr-atoms within the unit cell, and $\alpha$-phase gains stability at a relatively lower temperature. A detailed analysis reveals that in the $\alpha$-phase, there are displacements mainly in Sr1-atoms. At higher temperature around 650 $^0$C $\beta$-phase (\textit{R3c}) is stable. However, when the temperature is approaching the transition temperature,  T$_c$, the structural phonon softens, resulting in a phase transition around 150 $^0$C \cite{kalmutzki2017formation}. Therefore, low symmetry ($Cc$) $\alpha$-SrCNO becomes permanently polarized below 150 $^0$C. The spontaneous polarization for SrCNO and for the derivatives are listed in TABLE \ref{tab:3}.
%%%%%%%%%%%%%%%%%%%%%%%%%%%%%%%%%%%%%%%%%%%%%%%%%%%%%%%%%%%%%b%%%%%%%%%%%%%%%%%

The phonon structures are calculated with a dense K-point mesh of 8 $\times$ 8 $\times$ 8 through a supercell approach by utilizing the phonopy code as implemented in VASP\cite{phonopy-phono3py-JPCM}. As shown in Figure \ref{Figure2} (top panel), the absence of imaginary frequency in the low symmetry $\alpha$-phase dictates dynamical stability of the proposed ACNO ferroelectric MOFs. A similar conclusion can be drawn from the phonon total density of states as shown in Figure \ref{Figure2} (middle panel). Additionally, we find that MgCNO, CaCNO, SrCNO, and BaCNO show dominating peaks around 500 cm$^{-1}$, 800 cm$^{-1}$, 1000 cm$^{-1}$, and 300 cm$^{-1}$, respectively in the frequency scales. These dominating peaks might be important for thermal properties. We also investigate the element-resolved phonon density of states of SrCNO derivatives via Sr-site substitution with Mg, Ca, and Ba. We find that this substitution plays a pivotal role in exhibiting the largest peaks, as shown in Figure \ref{Figure2} (bottom panel). Together with a wide bandgap, their thermal properties can make these compounds important in the scientific community. However, it is worth noting that the current study focuses on structural, ferroelectric, and optical properties of the proposed compounds, which do not include thermal properties and will be addressed in the future.  

\textcolor{black}{To assess thermal stability, we additionally performed \textit{ab initio} molecular dynamics (AIMD) simulations for all $\alpha$-ACNO MOFs at 300 K, 500 K, and 700 K. The frameworks remain structurally intact up to 700 K, consistent with the experimental report that SrCNO is stable around 500 $^0C$ \cite{kalmutzki2017formation}. As expected, the average A–N and A–O bond lengths increase with temperature, while the overall framework connectivity is preserved (see Tables S1–S4 in the Supplementary Materials \cite{WB-FE-MOF}). These results collectively support the robustness of the ferroelectric phase and the feasibility of polarization switching in the proposed $\alpha$-ACNO MOFs.
}
%%%%%%%%%%%%%%%%%%%%%%%%%%%%%%%%%%%%%%%%%%%%%%%%%%%%%%%%%%%%%%%%%%%%%%%%%%%%%%
\begin{figure*} 
\centering
\includegraphics[width=0.8\linewidth]{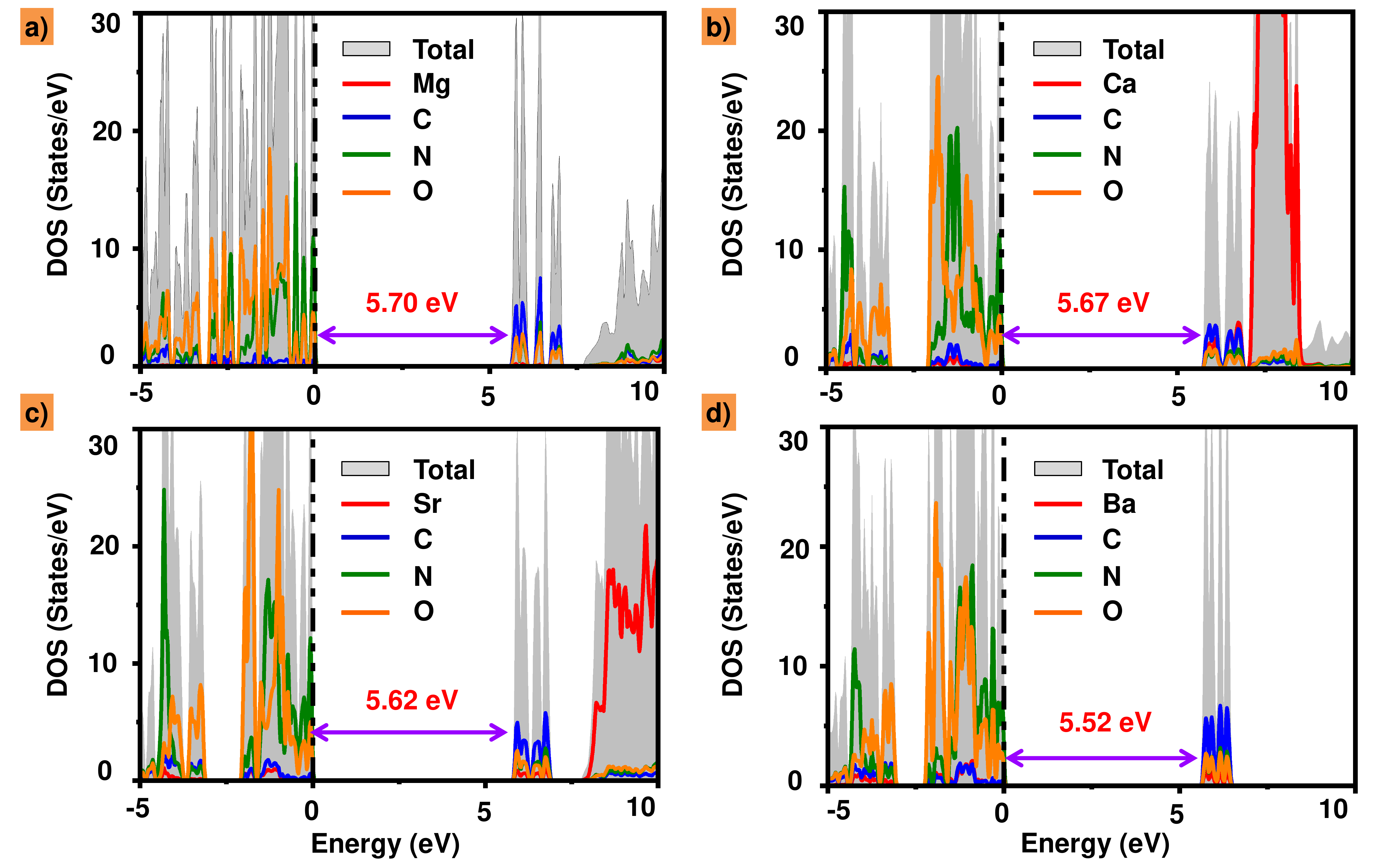}\vspace{-0pt}
\caption {\textbf{Electronic properties}. (a) MgCNO, (b) CaCNO, (c) SrCNO, and (d) BaCNO represent the total density of states for each atom in these systems. The results shown here are obtained using the GGA+MBJ function.} 
\label{Figure3}
\end{figure*}
%%%%%%%%%%%%%%%%%%%%%%%%%%%%%%%%%%%
\subsection{Electronic and optical properties of $\alpha$-ACNOs}
%%%%%%%%%%%%%%%%%%%%%%%%%%%%%%
Figure \ref{Figure3} depicts the calculated total density of states (DOS) of ACNO (A= Mg, Ca, Sr, Ba) system using the GGA+MBJ functional, respectively. The band gap (E$_g$) values are noted in TABLE \ref{tab:4} for all systems. As we can see, the E$_g$ of these materials is enhanced as the atomic size is decreased. The valence band is predominantly occupied by the hybridization of N-$2p$ and O-$2p$ states. whereas the conduction band is mainly driven by the overlapping of $2p$-states of C, and O atoms. In these materials, the optical transitions mainly can happen between overlapping of (N-$2p$, O-$2p$) and (C-$2p$, O-$2p$).     
%%%%%%%%%%%%%%%%%%%%%%%%%%%%%%%%%%%%%%%%%%%

\begin{figure*} 
\centering
\includegraphics[width=0.9\linewidth]{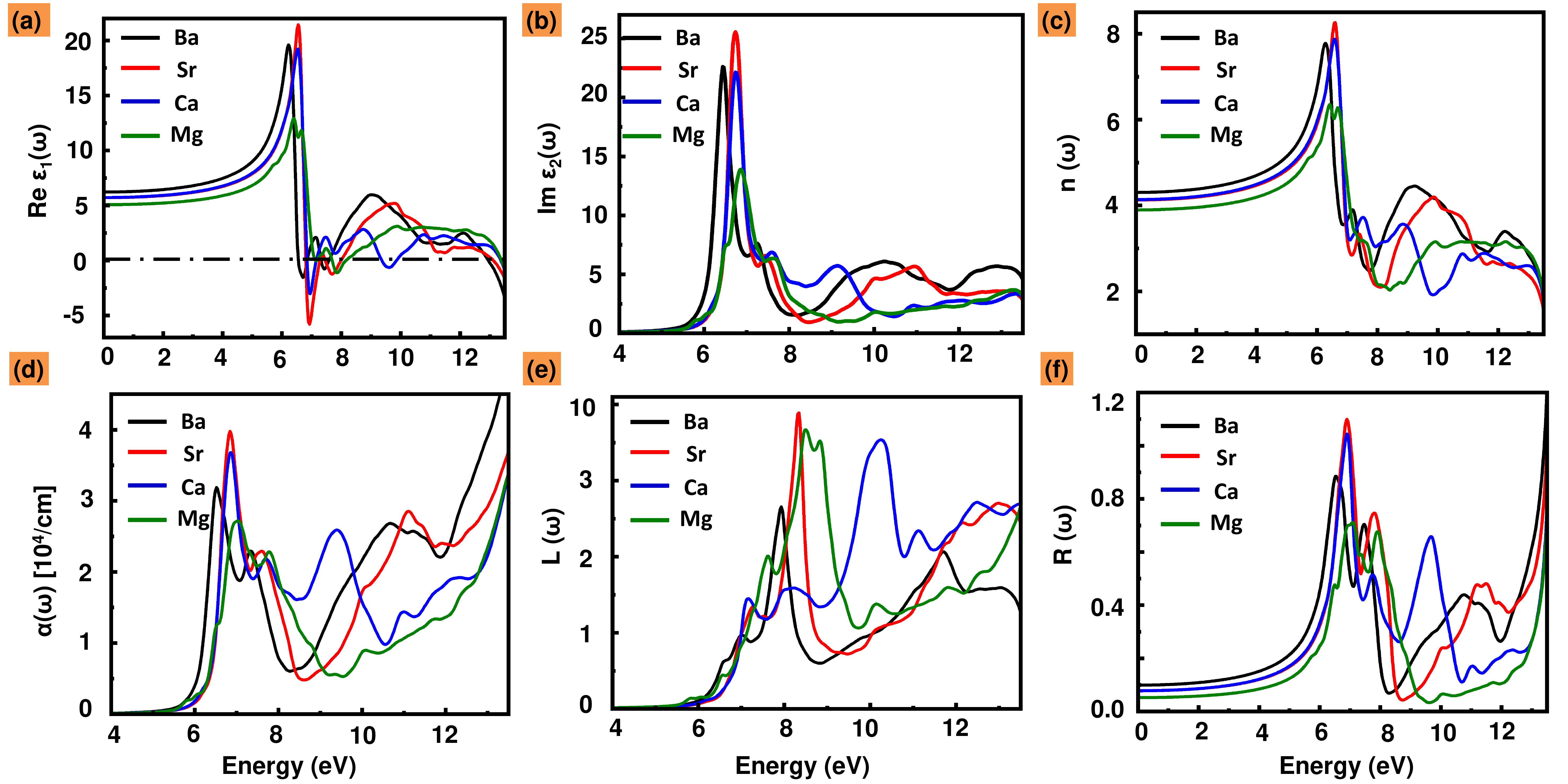}\vspace{-0pt}
\caption {\textbf{Optical properties}. The optical properties of BaCNO (Black), SrCNO (Red), CaCNO (Blue), and MgCNO (Green) systems. The figure represents (a) Real and (b) imaginary parts of the dielectric function, (c) refractive index, (d) absorption coefficient, (e) energy loss function, and (f) reflectivity
 with the variation of energy (eV). The results shown here are obtained using the GGA+MBJ calculation method.} 
\label{Figure4}
\end{figure*}

\textbf{Dielectric function}: The frequency-dependent complex dielectric functions are analyzed to determine the linear optical characteristics of matters \cite{ambrosch2006linear, wang2021vaspkit, blochl1994projector}. Imaginary components $\epsilon_2$($\omega$) of the dielectric function $\epsilon{(\omega)}$ are also computed from the electronic structures \cite{lucarini2005kramers}. Further, the Kramers-Kronig relations are used to calculate the real component $\epsilon_1$($\omega$) of the $\epsilon{(\omega)}$. Note that the total dielectric function is derived by the equation $\epsilon{(\omega)}$ =
$\epsilon_1{(\omega)}$ + $\epsilon_2{(\omega)}$. Using this total $\epsilon{(\omega)}$ we can calculate all other optical properties like refractive index $n$(0), absorption coefficient $\alpha$($\omega$), energy loss function $L$($\omega$) and reflectivity $R(\omega)$ \cite{john2017optical, saha2000electronic}. Figure \ref{Figure4}a represents the calculated $\epsilon_1{(\omega)}$ for four systems (color code represented in figures for each system). $\epsilon_1{(\omega)}$ gives the information about the material's dispersion and polarizability. The static dielectric constant $\epsilon_1{(0)}$ (at zero frequency) values are noted in Table \ref{tab:4} for all materials. Calculating the $\epsilon_1{(0)}$ value is important for ferroelectric materials. %The value of $\epsilon_1{(0)}$ is much lower in organic materials as compared to inorganic materials. 
%%%%%%%%%%%%%%%%%%%%%%%%%%%%%%%%
\begin{table}
\begin{center}
\setlength{\tabcolsep}{11pt}
\caption{\label{tab:4} \textbf{Optical properties.} Static dielectric function $\epsilon_1(0)$, static refractive index $n(0)$, static reflectivity $R(0)$, ferroelectric polarization \textbf{P}($\mu C/cm^2$), and energy band gap $E_g$(eV) for all compounds.}
\begin{tabular}{ccccc}
\hline
\hline
Compounds & $\epsilon_1(0)$ & $n(0)$ & $R(0)$ &  E$_g$(eV) \\
\hline
MgCNO     & 5.07  & 3.89 & 0.05 &  5.70   \\
CaCNO     & 5.71  & 4.12 & 0.07 &  5.67   \\
SrCNO     & 5.72  & 4.13 & 0.08 &  5.62   \\
BaCNO     & 6.22  & 4.30 & 0.09 &  5.52 \\ 
\hline
\hline
\end{tabular}
\end{center}
\end{table}
%%%%%%%%%%%%%%%%%%%%%%%%

$\epsilon_1{(0)}$ is inversely proportional to the binding energy of the electron-hole pairs. A large value of $\epsilon_1{(0)}$ will enhance the electron-hole pairs generation and increase the material's polarizability. Compared with four systems, BaCNO has a high $\epsilon_1{(0)}$ value. As a result, it has high polarization and low bandgap value (see Table \ref{tab:4}). After getting threshold energy (equal to the band gap) $\epsilon_1{(\omega)}$ for BaCNO, SrCNO, CaCNO and MgCNO spectra, it increases and gets the maximum value 19.60, 21.45, 19.23, and 12.92, respectively. After getting that sharp peak  $\epsilon_1{(\omega)}$ for all the compounds suddenly start to decrease and cross the zero value at 6.5, 6.84, 6.84 and 7.68 eV, respectively. This indicates the occurrence of plasma frequency. All the compounds show negative $\epsilon_1{(\omega)}$ value for a small energy region, thereafter $\epsilon_1{(\omega)}$ suddenly increases and gets a positive value. These negative and positive values of $\epsilon_1{(\omega)}$ indicate the metallic and semiconductor nature of the materials, respectively. In our systems, MgCNO has a high-energy range semiconductor character. After 13.0 eV energy point, all our compounds show a negative value, which indicates that after this energy point our materials permanently behave like a metal. Hereafter after our materials will only reflect the light.

%%%%%%%%%%%%%%%%%%%%%%%%%%%%%%%%%%%%%%%%%%%%%
Figure \ref{Figure4}b represents the $\epsilon_2{(\omega)}$ with variation of photon energy for all compounds. $\epsilon_2{(\omega)}$ is important to understand the absorption properties and energy loss mechanisms of materials. After getting threshold energy $\epsilon_2{(\omega)}$ gradually increasing and getting maximum value are 22.60, 25.55, 22.13 and 13.88 for ACNO (A= Ba, Sr, Ca, and Mg), respectively. Thereafter $\epsilon_2{(\omega)}$ suddenly decreases, indicating the absence of absorption in this energy region. All these compounds have one maximum peak in the UV light range; this peak arises mainly due to the transition from hybridization of Ni-$2s$ and O-$2s$ states to hybridization of C-$2p$ and O-$2p$ states. 

Figure \ref{Figure4}c represents the refractive index $n{(\omega)}$ of all our materials. The static refractive index $n{(0)}$ value for MgCNO, CaCNO, SrCNO, and BaCNO are 3.89, 4.12, 4.13, and 4.30, respectively. After getting band gap energy $n{(0)}$ start to gradually increasing and get maximum values are 6.27, 6.59, 7.87, and 6.35 for MgCNO, CaCNO, SrCNO, and BaCNO, respectively. The maximum value of $n{(0)}$ represents the maximum probability of an electron density energy point. The possibility of transition between the VB to CB can occur only in this energy point. All materials occurs maximum peak in UV light range. As a result, these materials can only absorb the UV light regions.

Figure \ref{Figure4}d shows the $\alpha$($\omega$) for all materials with variation of energy. when the materials get energy, which is equal to the band gap value $\alpha$($\omega$) start increasing and get maximum value of 318.84, 397.80, 368.13 and 271.89 × 10$^4$ cm$^{-1}$ for MgCNO, CaCNO, SrCNO and BaCNO respectively. All these materials have strong absorption in the UV light region. As a result, these materials can be used for optical applications in the UV light range. Figure \ref{Figure4}e shows the energy loss $L(\omega$) function with variation of energy in the unit of eV. After getting threshold energy $L(\omega$) starts increasing and reaches a maximum. After attaining plasma energy $L(\omega$) increases. Our materials have a minimum $L(\omega$) in the UV region. Figure \ref{Figure4}f shows energy-dependent reflectivity of four materials.  After obtaining plasma energy, all materials fully reflect the incident light (behave like metals). The findings of this investigation, all four materials absorb strongly in the UV light range. Substitution on the Sr-site slightly modifies the bandgap; yet these materials maintain their ultra-wide band gaps.

\section{Conclusion}
In summary, we investigate ferroelectric metal-organic frameworks using a combination of group theory techniques and $ab-initio$ DFT computations as potential candidates for wide-band-gap materials.
The key findings are:

\begin{itemize}
\item \textbf{Structural and ferroelectric properties}:

\begin{itemize}
 \item These ferroelectic metal-organic frameworks $\alpha$-A$_3$(O$_3$C$_3$N$_3$)$_2$ with A = Mg, Ca, Sr, and Ba crystallize in polar $Cc$-phase.

 \item The phase transition mechanism between non-polar to polar states is attributed to the displacement of Sr-atoms within the unit cell.

 \item The switching barrier for the bistable ferroelectric states is found to be smaller than that of other reported metal-organic frameworks.

 \item The spontaneous ferroelectric polarization values are found to be moderate in the range of 1.15 ($\mu C/cm^2$) to 2.13 ($\mu C/cm^2$).
\end{itemize}

\item \textbf{Electronic and optical properties}:
\begin{itemize}
\item These ferroelectric metal-organic frameworks, $\alpha$-A$_3$(O$_3$C$_3$N$_3$)$_2$ with A = Mg, Ca, Sr, and Ba, exhibit a wide-band-gap in the range of 5.5 eV to 5.7 eV that is crucial at high temperatures.

\item The possibility of utilizing them in the deep ultraviolet region is examined through absorption coefficients, dielectric constants measurements. We find that these compounds are potential candidates for ultra-wide band gap and hence deep ultraviolet region applications.
\end{itemize}

\item \textbf{Thermal and dynamical stabilities}:
\begin{itemize}
\item The $ab-initio$ molecular dynamics simulations show that the frameworks remain structurally intact up to 700 K. However, as expected, the average A–N and A–O bond lengths increase with temperature, while the overall framework connectivity is preserved.

\item Dynamic stability calculations suggest that these compounds are kinetically stable. The phonon analysis reveals that these ferroelectric MOFs (yet to be synthesized) can be grown in experiments.
\end{itemize} 
\end{itemize}

\section{Acknowledgment}
S.G. received research support from DST-SERB Core Research Grant File No. CRG/2018/001728.
Computations were performed using the High Performance Computing Cluster at SRM Institute of Science and Technology, Chennai, India.
%%%%%%%%%%%%%%%%%%%%%%%%%%%%%%%%%%%%%%%

\textbf{Competing interests}: The authors declare no competing interests.

\textbf{Data availability}: The data that support the results of this study are available from the authors upon reasonable request.
%%%%%%%%%%%%%%%%%%%%%%%%%%%%%%%%%%%%%%%%%%%%%%%%%%%%%%%%%%%%%%%%%%%%%

%%%%%%%%%%%%%%%%%%%%%%%%%%%%%%%%%%%%%%%%%%%%%%%%%%%%%%%%%%%%%%%%%%%%%
\bibliographystyle{apsrev4-2}

%\bibliography{myref}
%%%%%%%%%%%%%%%%%%%%%%%%%%%%%%%%%%%%%%%%%%%%%%%%%%%%%%%%%%%%%%%%%%%%%
\end{document}